\documentclass[conference]{IEEEtran}
\usepackage[latin1]{inputenc}
\inputencoding{latin1}

\usepackage{cite}
\ifCLASSINFOpdf
 \usepackage[pdftex]{graphicx}
\else
\fi
\begin{document}
%
% paper title
% Titles are generally capitalized except for words such as a, an, and, as,
% at, but, by, for, in, nor, of, on, or, the, to and up, which are usually
% not capitalized unless they are the first or last word of the title.
% Linebreaks \\ can be used within to get better formatting as desired.
% Do not put math or special symbols in the title.
\title{Experimental testbed for \\ seawater channel characterization}

% author names and affiliations
% use a multiple column layout for up to three different
% affiliations
\author{\IEEEauthorblockN{Pablo Mena, Pablo Dorta-Naranjo, \\ Gara Quintana, Iván Pérez-Álvarez \\ and Eugenio Jiménez }
\IEEEauthorblockA{Instituto para el Desarrollo Tecnológico \\y la Innovación en Comunicaciones (IDeTIC),\\
Universidad de Las Palmas de Gran Canaria,\\
Las Palmas, 35017, Spain.\\
Email: pmena@idetic.eu}
\and
\IEEEauthorblockN{Santiago Zazo \\ and Marina Pérez}
\IEEEauthorblockA{ETS Ingenieros de Telecomunicación\\
Universidad Politécnica de Madrid (UPM),\\ Madrid 28040, Spain \\
Email: santiago@gaps.ssr.upm.es}
\and
\IEEEauthorblockN{Laura Cardona \\ and J. Joaquín Hernández}
\IEEEauthorblockA{Plataforma Oceánica \\ de Canarias (PLOCAN)\\
Telde 35200 (Spain)\\
Email: laura.cardona@plocan.eu}}

% conference papers do not typically use \thanks and this command
% is locked out in conference mode. If really needed, such as for
% the acknowledgment of grants, issue a \IEEEoverridecommandlockouts
% after \documentclass

% for over three affiliations, or if they all won't fit within the width
% of the page, use this alternative format:
% 
%\author{\IEEEauthorblockN{Michael Shell\IEEEauthorrefmark{1},
%Homer Simpson\IEEEauthorrefmark{2},
%James Kirk\IEEEauthorrefmark{3}, 
%Montgomery Scott\IEEEauthorrefmark{3} and
%Eldon Tyrell\IEEEauthorrefmark{4}}
%\IEEEauthorblockA{\IEEEauthorrefmark{1}School of Electrical and Computer Engineering\\
%Georgia Institute of Technology,
%Atlanta, Georgia 30332--0250\\ Email: see http://www.michaelshell.org/contact.html}
%\IEEEauthorblockA{\IEEEauthorrefmark{2}Twentieth Century Fox, Springfield, USA\\
%Email: homer@thesimpsons.com}
%\IEEEauthorblockA{\IEEEauthorrefmark{3}Starfleet Academy, San Francisco, California 96678-2391\\
%Telephone: (800) 555--1212, Fax: (888) 555--1212}
%\IEEEauthorblockA{\IEEEauthorrefmark{4}Tyrell Inc., 123 Replicant Street, Los Angeles, California 90210--4321}}

% use for special paper notices
%\IEEEspecialpapernotice{(Invited Paper)}

% make the title area
\maketitle

% As a general rule, do not put math, special symbols or citations
% in the abstract
\begin{abstract}
%Most underwater communications use acoustic or optical methods for wireless transmissions. Electromagnetic (EM) communications have been disregarded due to attenuation at high frequencies. 
Shallow seawaters are problematic for acoustic and optical communications. Sensor networks based on electromagnetic (EM) communications are evaluated in this environment. In order to characterize the subaquatic channel, several measurement systems have been designed, built and tested in the sea obtaining very reliable results. Experiments carried out with dipoles and loop antennas showed serious disagreement with the state of the art, especially when dipole antennas are used. Dipoles performance was poor while magnetic loops showed relevant results. Measurement system is described in detail and real attenuation of the subaquatic channel is obtained for several distances and antennas. Finally, measured and simulated results are compared with good agreement.
\end{abstract}

% no keywords

% For peer review papers, you can put extra information on the cover
% page as needed:
% \ifCLASSOPTIONpeerreview
% \begin{center} \bfseries EDICS Category: 3-BBND \end{center}
% \fi
%
% For peerreview papers, this IEEEtran command inserts a page break and
% creates the second title. It will be ignored for other modes.
\IEEEpeerreviewmaketitle

\section{Introduction}
It is generally accepted that acoustic communications are unbeatable in deep sea water when no power limitations or high data rates are required. However, it is in shallow waters where acoustic systems show much worse performance\cite{che2010re}\cite{uribe2009radio}.
Underwater acoustic suffers from its low speed (1500 m/s) while the restricted operating frequencies limit the bandwidth and therefore the data rate. In shallow seawater or ports acoustic systems are interfered by multiple sources of noise like industries or boats nearby. In other words, the effectiveness of propagation of acoustic signals is inevitably accompanied by the unwanted effects of multi-path, reverberation, ambient noise and a negative impact on marine life\cite{parsons2008navy}\cite{jepson2003gas}. Similarly, free space optical communications are impractical in shallow waters because the severe water absorption and strong backscatter from suspending particles\cite{wang2014experimental}. Even the clearest water has 1000 times the attenuation of clear air, and turbid water has 100 times the attenuation of the densest fog\cite{lanbo2008prospects}. 

According to some old times perspectives,  EM communications have been overlooked due to heavy attenuation at high frequencies. However, they overcome all the negative aspects that acoustic and optical systems suffer in shallow seawater. Many studies dealing with EM signal in seawater describe alternative propagation mechanisms\cite{pelavas2014development}\cite{wang2014experimental}. The shallow water environment presents two interfaces: water to air, and water to seafloor (both seafloor and air with less conductivity than seawater). According to literature, if the communication takes place near the interface with a less conducting medium, the performance of EM communication is actually better than what can be predicted in a homogeneous space. These mechanisms are called by different names like 'up-over-down propagation' or 'lateral wave propagation'. Some experiments with submerged electric or magnetic antennas have been conducted to verify this propagation mechanisms with more or less success\cite{cella2009electromagnetic}. The technical difficulty of these measures (specially in real scenarios) is not sufficiently studied. It is common to obtain apparently valid results just to find lack of reproducibility or coherence. Thus in addition to discussing different antenna configurations and propagation experiments, this paper is directed towards illuminating the practical problem of conducting high sensitivity measurements in a real water environment. Several measurements systems are described and debugged until a reliable system is obtained. Finally, results showed significant conflicts with some publications while other works are verified. 

	 Measurements were performed in a controlled environment and  in the sea to confirm experimentally propagation mechanisms described in literature and simulated numerical solutions. Nevertheless, some early results were discarded based on coherence and basic data analysis.  It has been a hard and long procedure to detect all interferences and bad measurement procedures. To perform these experiments a signal generator and a high sensitive time and frequency domain instruments were required. All instruments had to be capable of being controlled remotely and save digital data on demand.
	
\section{Water tank preliminaries}

\begin{figure}[!!!!ht]
		\begin{center}
			\includegraphics[width=9cm]{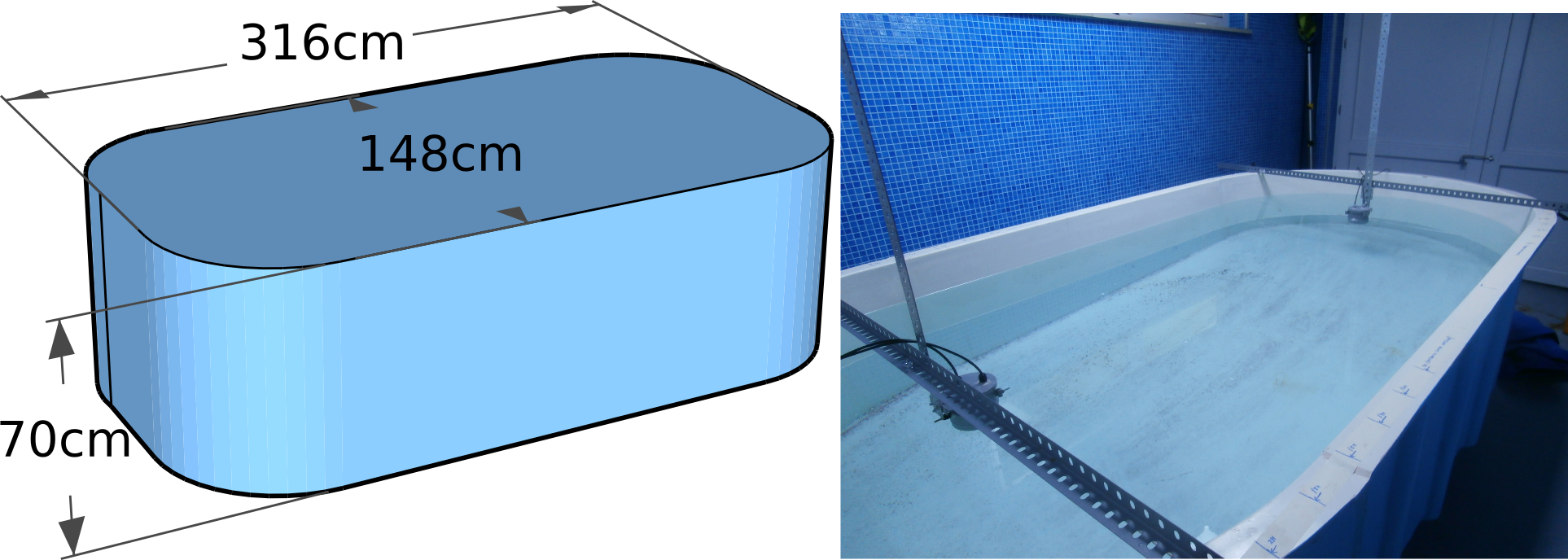}
		\end{center}
		\caption{Water tank dimensions.}
		\label{fig:watertank}
\end{figure}

\begin{figure}[!!!!ht]
		\begin{center}
			\includegraphics[width=7cm]{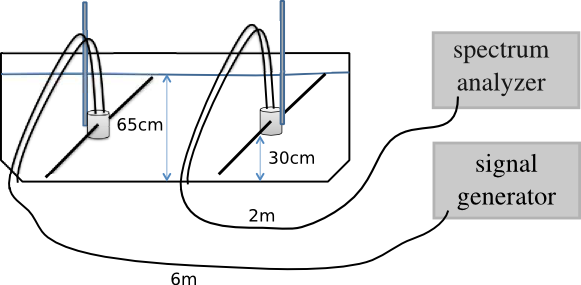}
		\end{center}
		\caption{Controlled indoors experiment.}
		\label{fig:diagramTank}
\end{figure}

A controlled environment was set in PLOCAN facilities where a sizable fiber tank was filled with seawater from nearby shore. This marine research institute is located conveniently close to the harbour where sea experiments were conducted. The seawater tank (Fig. \ref{fig:watertank})
 was useful making preliminary experiments with a basic testbed. In addition, all physical aspects of water seals were evaluated before being placed in the  sea. In this first approach only antennas were placed in the water connected to instruments by RG-58 coaxial cables of 50$\Omega$ impedance. Time domain measurements were made with a Digital Oscilloscope (Agilent MSO6032) while a Spectrum Analyzer (Agilent E4404B) was used for low power frequency domain analysis. Testbed is showed in Fig. \ref{fig:diagramTank}, the antennas were at 30cm from the bottom of the tank suspended by vertical bars. Two coaxial cables of 2 and 6 meters were used to connect receiver and transmitter respectively.
 %Two antenna types were used in the water tank: single dipole and crossed dipole. 

 	Dipole elements were formed by 59 cm long stainless round bars with a diameter of 55 mm. Each bar was screwed to a sealed PVC receptacle. At first, self-vulcanizing tape was used to seal the bars but total water sealing was difficult to achieve at bar joints. A second version of these antennas, developed during field measurements, solved this problem: thin PVC pipes around the metallic bars provide insulation and joints with the PVC receptacle were sealed with a screwed pipe coupling.
	Three kinds of dipoles were built based on \cite{cella2009electromagnetic} \cite{al2004propagation}: insulated, uninsulated terminations and totally uninsulated. In the uninsulated configuration dipole bars were in direct contact with water while the uninsulated terminations were 2 cm long in the tip of each dipole element. Inside the water seal, baluns were used to provide a balanced excitation to each antenna since coaxial is an unbalanced transmission line (with one terminal at ground potential). Electrical imbalance could cause currents to circulate on the outer shields of coaxial lines providing a mechanism for extraneous fields to leak into the receiver.

	Attenuation values were obtained for three frequencies (100, 200 and 300 KHz) and different separations (tank length limited distances to 1-2 m). The signal generator fed dipoles with a tone of 20 dBm. Crossed dipoles were fed in phase quadrature with 17 dBm each to maintain the same power. Signal power at each frequency was measured with a spectrum analyzer. Even taking into account reduced light speed in seawater, dipoles lengths made them inefficient at these frequencies since wavelengths are in the order of thousands of meters. Anyway, they can be considered as equally inefficient for all the frequencies studied. This is the reason why they are extremely useful, their frequency response does not affect  broadband measurements to characterize the medium. In order to evaluate measures, simulations were carried out using FEM (Finite Element Method) on a 3D model of the experiment.  
	
\begin{figure}[!!ht]
		\begin{center}
			\includegraphics[width=8.5cm]{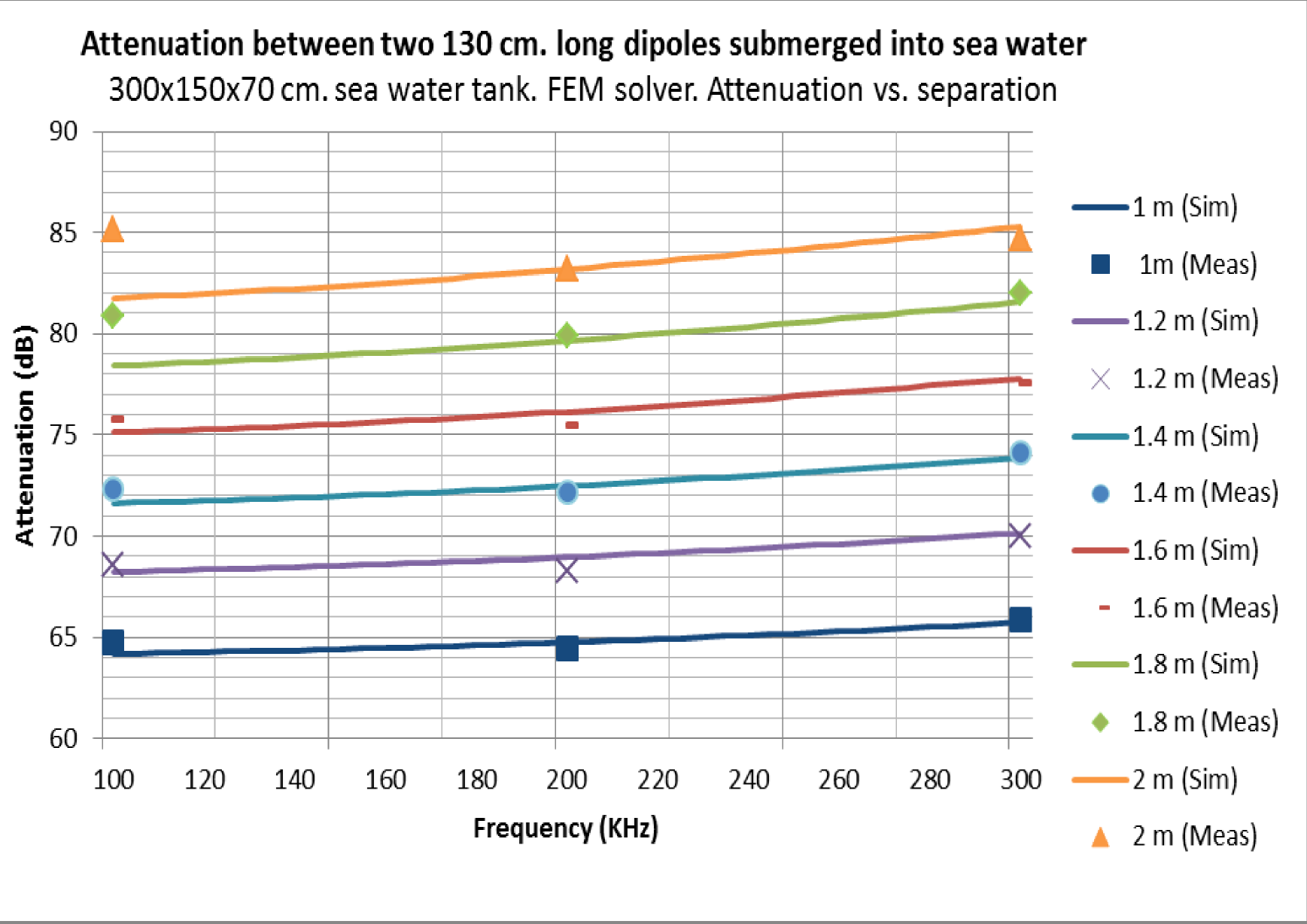}
		\end{center}
		\caption{Attenuation between two 130cm long uninsulated dipoles submerged into seawater tank. Simulated and measured results.}
		\label{fig:graphTank}
\end{figure}

	This preliminary experiment was far from giving useful measures due to tank limited size, but it was necessary to test equipment, water seals and antennas.  Despite this fact, a good agreement between simulated and measured results was obtained. (Fig. \ref{fig:graphTank})
		
\section{Field measurements}

\begin{figure}[!!ht]
		\begin{center}
			\includegraphics[width=7cm]{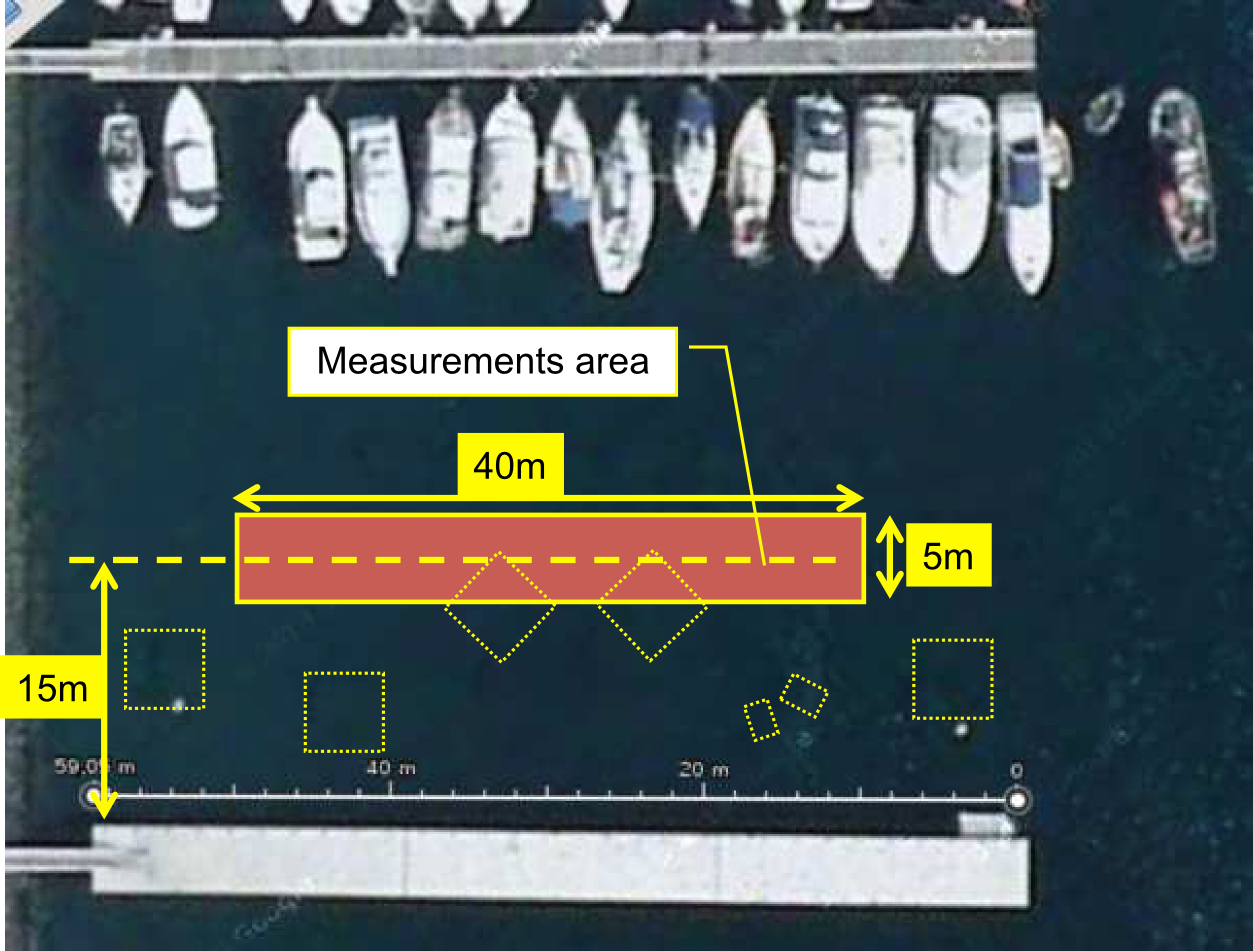}
		\end{center}
		\caption{Measurement area besides the floating pier.}
		\label{fig:measArea}
\end{figure}

\begin{figure}[!!ht]
		\begin{center}
			\includegraphics[width=7cm]{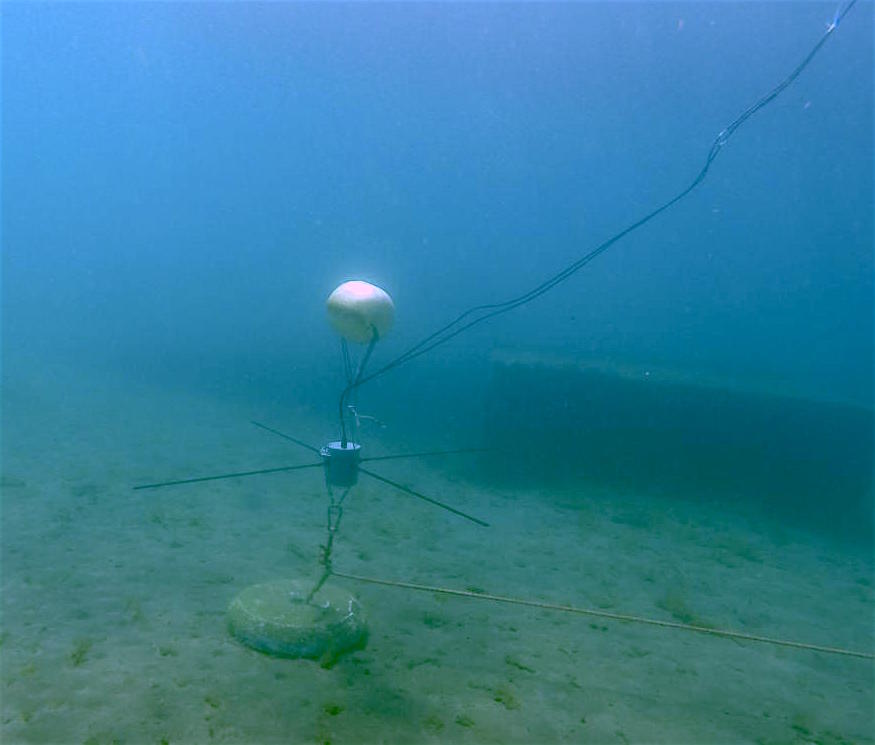}
		\end{center}
		\caption{Measurement setup with instruments outside the water.}
		\label{fig:measSetup}
\end{figure}

At field site, experiments were located between 30 to 50 meters off the shore line and parallel to a concrete floating pier (Fig. \ref{fig:measArea}) in PLOCAN facilities. Monitoring was performed on the pier, 15 meters separated from experiments area while receiver and transmitter were placed at several distances along the pier side. A pulley system with a buoy and a heavy weight allowed different heights for the antennas while different distances were setup by a diver manually. When different heights were needed, each antenna was hold in its position by the tension between a submerged buoy and a pulley attached to a heavy weight (Fig. \ref{fig:measSetup}). The rope passed through the pulley and was tied to the pier to control height from the surface. %When measurements where made on the seabed, antennas were situated directly on the sand since weight and pulley could disturb the results.

At first, instruments were located on the pier with 30m long RG-58 coaxial cables between the surface and the antennas. More than 15 measurement campaigns were carried out along 2015 and 2016. However, severe aerial coupling between cables with this arrangement made impossible to obtain reliable measurements. Signal leakage is not negligible in RG-58 cable for this frequency range \cite{173352} since extremely weak signals are measured. Moreover, coupling between coaxial cables is higher at low frequencies and long cables. It was determined that signal was induced mainly through air since the more cable was inside the water the less coupling occurred. To reduce cable lengths as much as possible the instruments must be located at a short distance from the antennas. As a consequence, waterproof vessels for instruments were designed and built. This reduced coupled signals to undetectable levels making possible to obtain seawater noise measurements very similar to the default analyzer noise floor spectrum (with all signal inputs disconnected). As a matter of fact, seawater environment was found to have an extremely low noise electromagnetically speaking. 

\subsection{Second version: submerged instruments}

\begin{figure}[!!ht]
		\begin{center}
			\includegraphics[width=9cm]{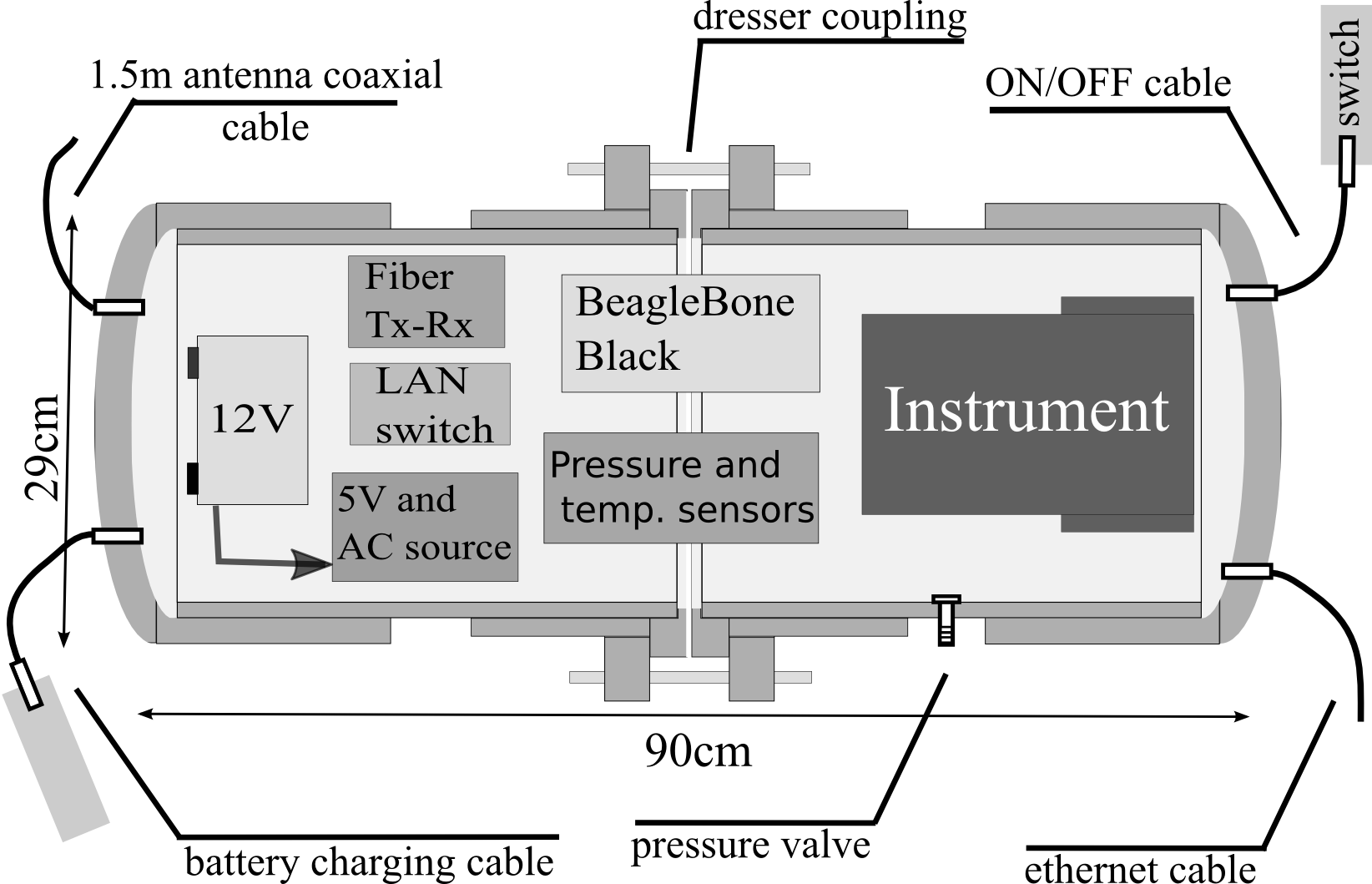}
		\end{center}
		\caption{Pressure vessel with approximated location of electronics.}
		\label{fig:cilinder}
\end{figure}

Waterproof containers for the electronics consisted of large diameter pipes with a dresser coupling in the middle and two caps on each side. They were designed to enclose a waveform generator (Keysight 33220A) or the spectrum analyzer (Agilent N9340B), labeled as \emph{Instrument} in Fig. \ref{fig:cilinder}. All cables had their own plastic covering and were passed through the pipe by cable glands. When the fitting was screwed down, sealing grommets around the cable provide a watertight seal. To ensure water sealing the interior of the vessel was pressurized as seen in \cite{siegel1973electromagnetic}. To accomplish this, air valves were installed on the sides of both vessels. If sufficient air pressure was stored it would prevent water leaking inside the vessel if water seal failed. Instruments were monitored from surface using their built-in LAN capability and Keysight VEE software. Inside, a BeagleBone Black board was connected to the same LAN through a switch to monitor temperature, pressure and power on/off. All electronics were powered with a battery inside the vessel since an external power source might produce the same interference problems. A sealed battery  (12V 30A/h) provided power to instruments and electronics. Some instruments needed AC power while others needed 5 or 12V so a power inverter and DC-DC voltage converter were installed. All these electronics were tested and filtered as required so they would not affect measurements. The only cables set between each vessel and surface were used for ethernet communications.

	Ethernet physical connection was established using shielded twisted pair (STP) 30m cables but it was found as problematic as coaxial. In this case, signal travelled from signal generator LAN port to the surface and some of this signal was induced in the other cable. That other cable was connected to the spectrum analyzer LAN port were this interference was measured. This alternative signal path unvalidated water-to-water measurements since interference level was not negligible. Coupling problems were solved completely using optical fiber to communicate water and surface. For this purpose, two Ethernet-to-Fiber media converters were installed inside both vessels and STP cables were replaced by two 30 meter flexible fiber cables.
	
	Measurement procedure typically started recording noise received by antennas and probe cable with the spectrum analyzer and signal generator disconnected. Those measurements were carried out on the pier testing the equipment on air and then submerged to find interferences.
		
\begin{figure}[!!ht]
		\begin{center}
			\includegraphics[width=7cm]{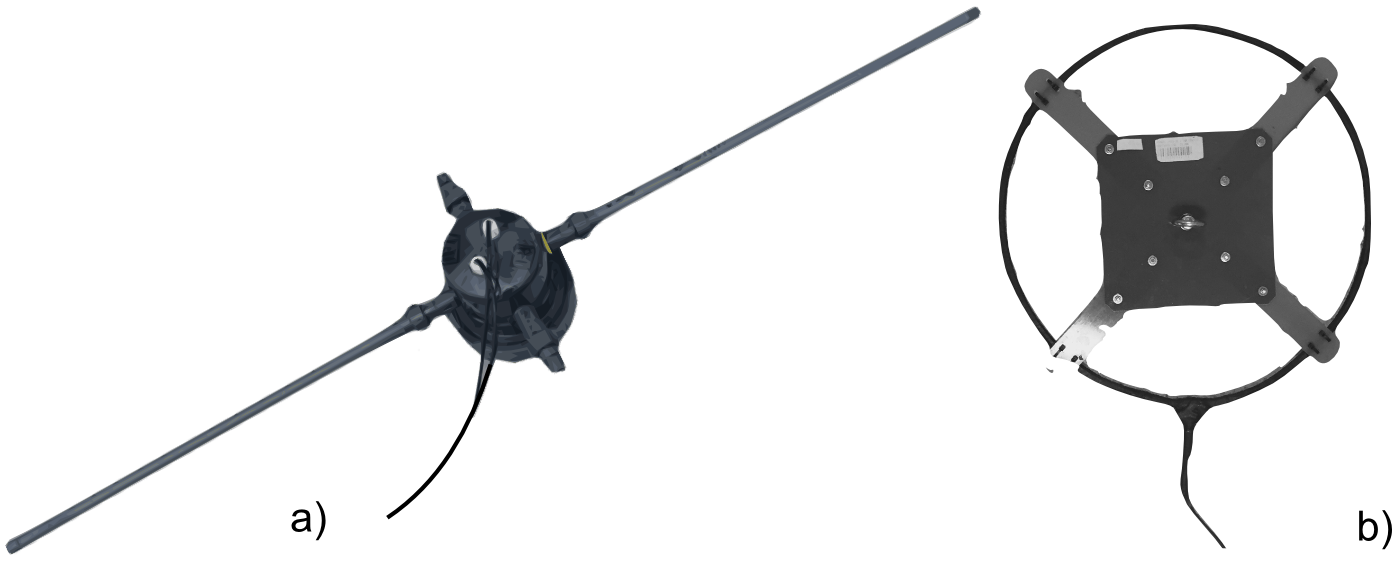}
		\end{center}
		\caption{Used antennas. (a) Dipole (insulated configuration). (b) 10-turns Loop. }
		\label{fig:antennas}
\end{figure}

	 At this point, attenuation with several antenna configurations was measured on seabed at several distances to test the system. Insulated dipoles were found to be too extremelly inefficient, even at 1 meter of distance attenuation was above 140 dB. For this reason, insulated dipoles \ref{fig:antennas}(a) were discarded for further tests. While writing this document new measurement campaigns have been scheduled since these preliminary results contradict several works \cite{al2004propagation}\cite{cella2009electromagnetic} previously published in the literature. 
	 
	 On the other hand, loop antennas beat dipoles performance in all situations studied. Loop antennas consist on ten turns of copper magnet wire forming a 30cm diameter circumference as seen in Fig. \ref{fig:antennas}(b). Each wire had a thin layer of insulating coating and the whole coil was covered with self-vulcanizing tape. The antenna was mounted on a glass fiberboard plate to provide rigidity and had a built-in sealed balun where coaxial cable connection was made.

\section{Results}

\begin{figure}[!!ht]
		\begin{center}
			\includegraphics[width=9cm]{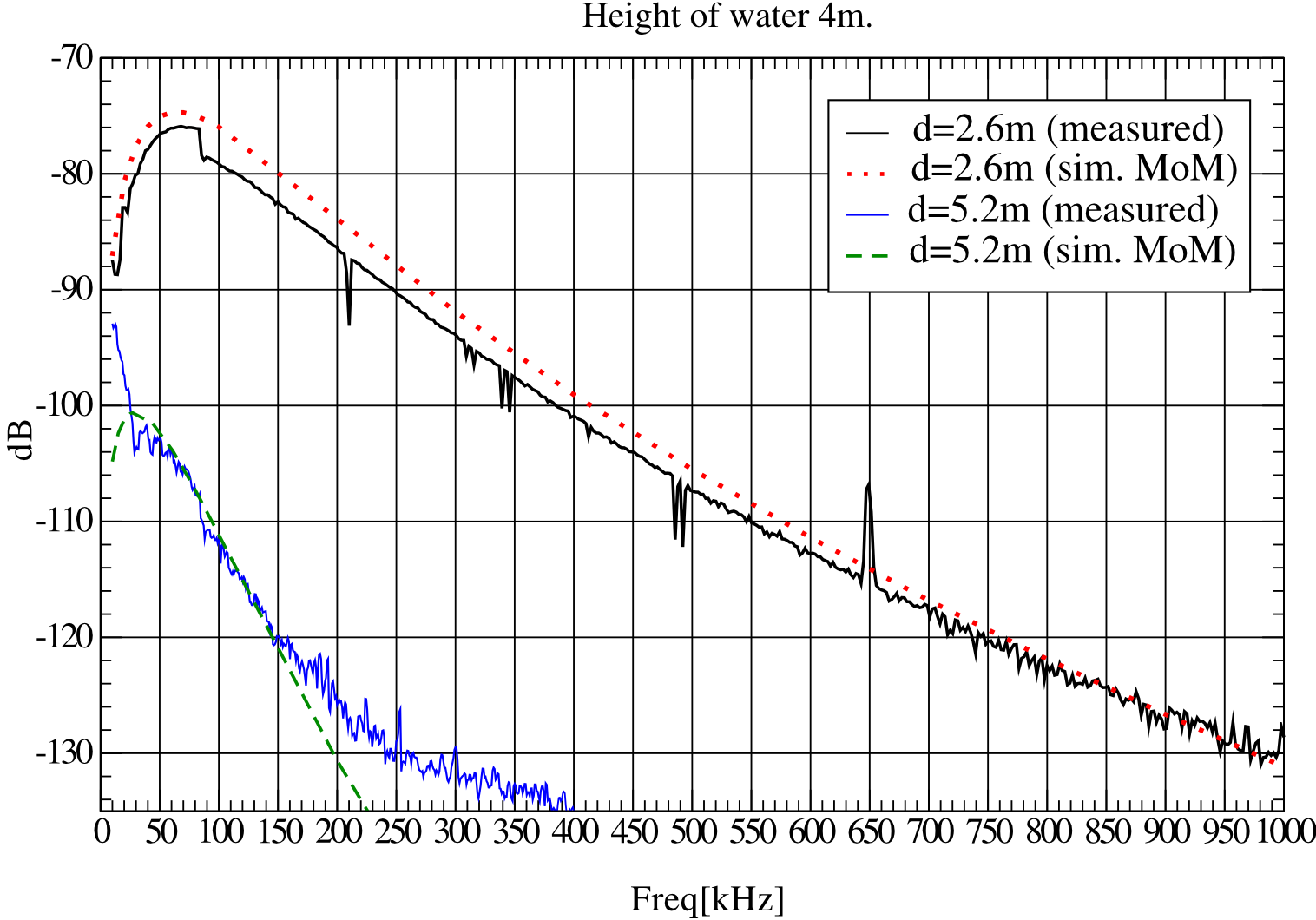}
		\end{center}
		\caption{Attenuation between two horizontal ten-turns loops placed on seabed. Simulated and measured results. }
		\label{fig:graphFinal}
\end{figure}

Once a reliable testbed was built and tested, basic measurements were carried out to confirm simulations accuracy. As seen in the Fig. \ref{fig:graphFinal} frequency sweeping was done between 10 KHz and 1 MHz for two distances (2.6 and 5.2 meters). In these measurements we used the loop antennas placed directly on seabed horizontally while distance between their centres was set manually by a diver. In order to obtain the frequency response of the channel, the spectrum analyzer maximum hold feature was set while a tone sweeping was slowly detected on each frequency. Therefore, attenuation was calculated based on transmitted power (18 dBm). Simulations were performed using a commercial MoM (Method of Moments) solver: FEKO. Despite of spectrum analyzer was not suited for low frequencies, simulations and measurements showed good agreement. Calibration curves were obtained experimentally to improve its response. A full study with further measurements, comparison with simulations and analysis of the results were carried out with the same testbed in \cite{eugenioUcomms16}.

\begin{figure}[!!ht]
		\begin{center}
			\includegraphics[width=9cm]{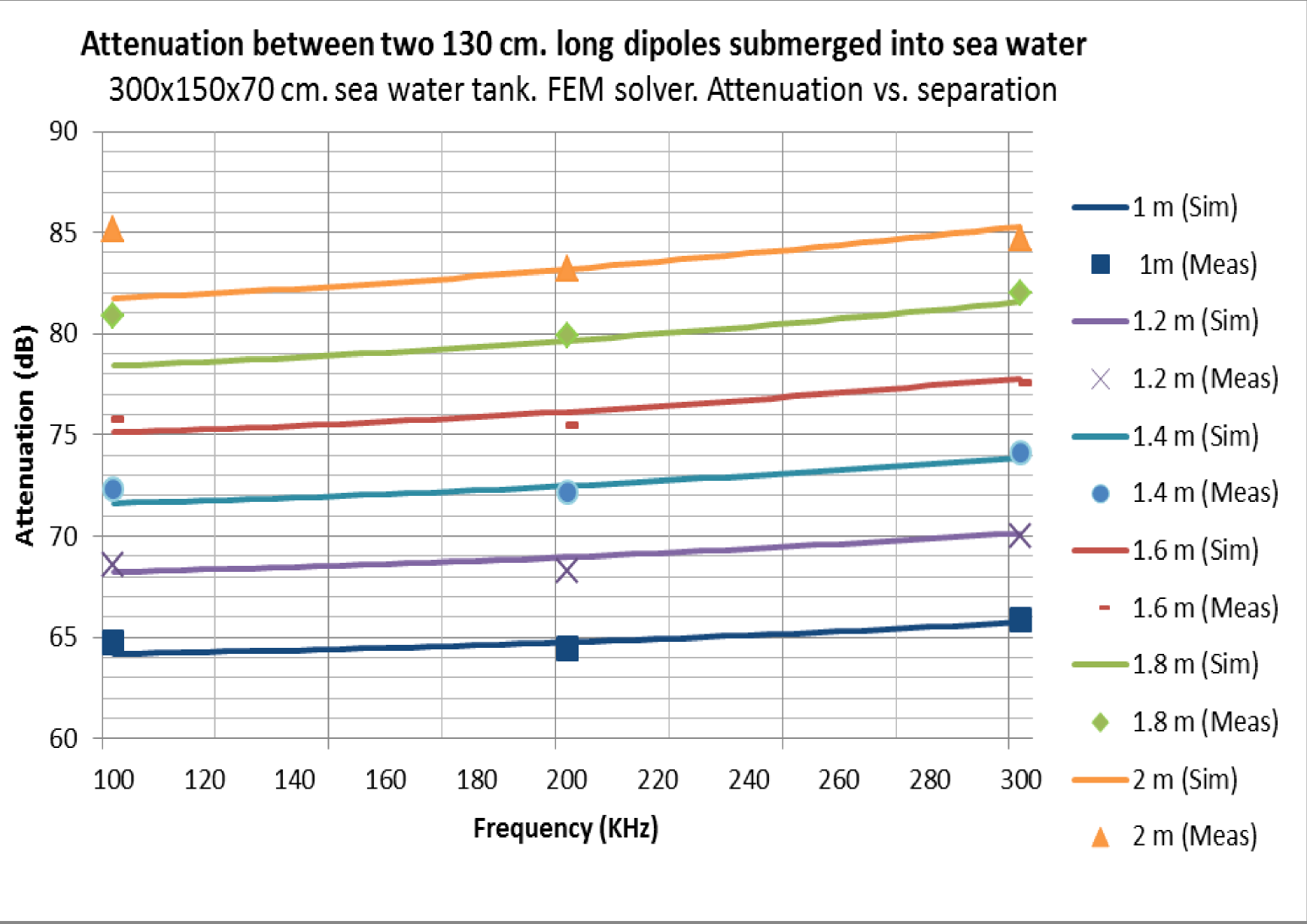}
		\end{center}
		\caption{Amplitude time evolution for TX-RX = 4m at 90 KHz. }
		\label{fig:graphTime}
\end{figure}

Time variation is an important parameter for a positioning system based on power received\cite{jzazoUcomms16}. Evolution of signal strength at 90 KHz within 2 hours is showed in Fig. \ref{fig:graphTime}. Time stability was found to be very high for RF communication standards. A very slow long-term variation is observed while in the short-term it showed a noise variance of $\pm0.1$ dB probably due to instruments amplitude accuracy\cite{N9340TechReport}. 

\section{Conclusion}

Coupling of signals between cables was found to be a serious problem in underwater channel measurements. Different testbeds were carried out and debugged during a number of measurement campaigns. A final testbed is compared with simulations showing high agreement. This coherence with theory validates our testbed and allows us to use simulated results confidently.

\section*{Acknowledgment}

The authors would like to thank the work carried out by Juan Domingo Santana Urbín (ULPGC) when making the pressurized PVC receptacles, loop antennas and all of the lab stuff needed to make the measurements. Also, the authors would like to thank the work carried out by Gabriel Juanes and Raul Santana (PLOCAN) when setting up measurement testbed in the pier and into the sea.
This work has been supported by Ministerio de Economía y Competitividad, Spain, under public contract TEC2013-46011-C3-R.

% trigger a \newpage just before the given reference
% number - used to balance the columns on the last page
% adjust value as needed - may need to be readjusted if
% the document is modified later
%\IEEEtriggeratref{8}
% The "triggered" command can be changed if desired:
%\IEEEtriggercmd{\enlargethispage{-5in}}

% references section

% can use a bibliography generated by BibTeX as a .bbl file
% BibTeX documentation can be easily obtained at:
% http://mirror.ctan.org/biblio/bibtex/contrib/doc/
% The IEEEtran BibTeX style support page is at:
% http://www.michaelshell.org/tex/ieeetran/bibtex/
\bibliographystyle{IEEEtran}
% argument is your BibTeX string definitions and bibliography database(s)
%\bibliography{IEEEabrv,../bib/paper}
%
% <OR> manually copy in the resultant .bbl file
% set second argument of \begin to the number of references
% (used to reserve space for the reference number labels box)
\bibliography{bibliography}

% that's all folks
\end{document}